\documentclass{aa}
\usepackage{txfonts} 
\usepackage{graphicx}
\usepackage[colorlinks=true, linkcolor=blue, citecolor=blue, urlcolor=blue]{hyperref}
\usepackage{natbib}

\newcommand{\halpha}{H\,$\alpha\ $}
\newcommand{\heII}{He\,{\sc ii} ($\lambda$4686 \AA) }

\begin{document}
  \title
    {Exploring inside-out Doppler tomography:
     non-magnetic cataclysmic variables}
  \author
    {E.~J.~Kotze\inst{\ref{inst1},\ref{inst2}}
     \and
     S.~B.~Potter\inst{\ref{inst1}}
     \and
     V.~A.~McBride\inst{\ref{inst1},\ref{inst2}}}
  \institute
    {South African Astronomical Observatory,
     PO Box 9, Observatory 7935, Cape Town, South Africa\\
     \email{ejk@saao.ac.za}\label{inst1}
     \and
     Astrophysics, Cosmology and Gravity Centre (ACGC),
     Department of Astronomy, University of Cape Town,
     Private Bag X3, Rondebosch 7701, South Africa\label{inst2}}
  \date
    {Received 22 April 2015 / Accepted 25 May 2015}
  \abstract
    {Doppler tomography is a technique that has revolutionised the
     interpretation of the phase-resolved spectroscopic observations of
     interacting binary systems.}
    {We present the results of our investigation of reversing the velocity axis
     to create an inside-out Doppler coordinate framework with the intent to
     expose overly compacted and enhance washed out emission details in the
     standard Doppler framework.}
    {The inside-out tomogram is constructed independently of the standard
     tomogram by directly projecting phase-resolved spectra onto an inside-out
     velocity coordinate frame.
     For the inside-out framework, the zero-velocity origin is transposed to the
     outer circumference and the maximum velocities to the origin of the
     velocity space.
     We test the technique on a simulated system and two real systems with
     easily identifiable features, namely the accretion disc and bright spot in
     WZ Sge, and spiral shocks in IP Peg.}
    {Our tests show that there is a redistribution of the relative brightness of
     emission components throughout the tomograms, i.e., where the standard
     framework tends to concentrate and enhance lower velocity features towards
     the origin, the inside-out velocity framework tends to concentrate and
     enhance higher velocity features towards the origin.
     Conversely, the standard framework disperses and smears the higher
     velocities farther away from the origin whereas the inside-out framework
     disperses and smears the lower velocities.
     In addition, the projection of the accretion disc in velocity space now
     appears correctly orientated with the inner edge close to the maximum
     velocity origin and its outer edge closer to the zero velocity outer
     circumference.
     Furthermore, the gas stream and secondary star are projected on the outside
     of the disc with the bright spot of the stream-disc impact region on the
     disc's outer edge in the inside-out velocity space.
    }
    {We conclude that inside-out Doppler tomography complements the already
     powerful tomographic techniques in the analysis of spectroscopic emissions
     from interacting binary systems.}
  \titlerunning
    {Exploring inside-out Doppler tomography}
  \authorrunning
    {E.~J.~Kotze et al.}
  \maketitle

\section{Introduction}
\label{sec:Intro}

  Cataclysmic variables (CVs) are interacting binary systems in which a white
  dwarf (primary) accretes material from a late-type main-sequence star
  (secondary).
  The secondary fills its Roche lobe and material flows through the inner
  Lagrangian point ($L_{1}$) towards the primary.
  The material may form an accretion disc around the primary before it is
  finally accreted
  \citep[see][for a comprehensive review of CVs]{1995CAS....28.....W}.

  \citet{1988MNRAS.235..269M}
  developed Doppler tomography as a technique
  aimed at constructing a two-dimensional velocity image (tomogram) of the
  accretion discs of CVs using an emission line in its spectra sampled at a
  number of orbital phases.
  Doppler tomography has revolutionised the interpretation of orbitally
  phase-resolved spectroscopic observations of interacting binary systems.
  It has become a valuable tool for resolving the distribution of line emission
  in CVs and other binary systems
  \citep[see, e.g.,][Astrotomography workshop review]{2001LNP...573....1M}.
  Significant contributions have been, amongst others, the discovery of two-arm
  spiral shocks in the accretion disc of the dwarf nova \object{IP Peg}
  \citep{1997MNRAS.290L..28S}
  and mapping of the accretion stream in the polar \object{HU Aqr} detailing the
  transition from ballistic to magnetic flow
  \citep{1997A&A...319..894S}.
  Recent literature also abounds with examples where emission from the secondary
  and the bright spot of the stream-disc impact region in CVs have been resolved
  in Doppler tomograms, such as \object{CTCV J1300-3052}
  \citep{2012MNRAS.422..469S}
  and \object{V455 And}
  \citep{2013MNRAS.429.3433B}.

  We present a complementary technique for an inside-out velocity projection for
  Doppler tomography.
  This paper expands on the preliminary results presented in
  \citet{2015KotzePotter}.
  In Section \ref{sec:VelSpace} we first review the cartesian coordinate frame
  used in Doppler tomography and introduce a polar coordinate frame for the new
  inside-out velocity space.
  Next, in Section \ref{sec:DopTom} we compare the standard and inside-out
  Doppler tomograms for a simulated disc-accreting system and two real CV
  systems, \object{WZ Sge} and IP Peg.
  Also in Section \ref{sec:DopTom}, we investigate how the inside-out framework
  redistributes relative contrast levels amongst the various emitting components
  of observations of a variety of CV features.
  Finally, in Section \ref{sec:Summary} we provide a summary and conclusions.

\section{Velocity space}
\label{sec:VelSpace}

\subsection{Velocity in cartesian coordinates}
\label{sec:VelCart}
  \begin{figure}
  \centering
  \includegraphics[width=8cm]{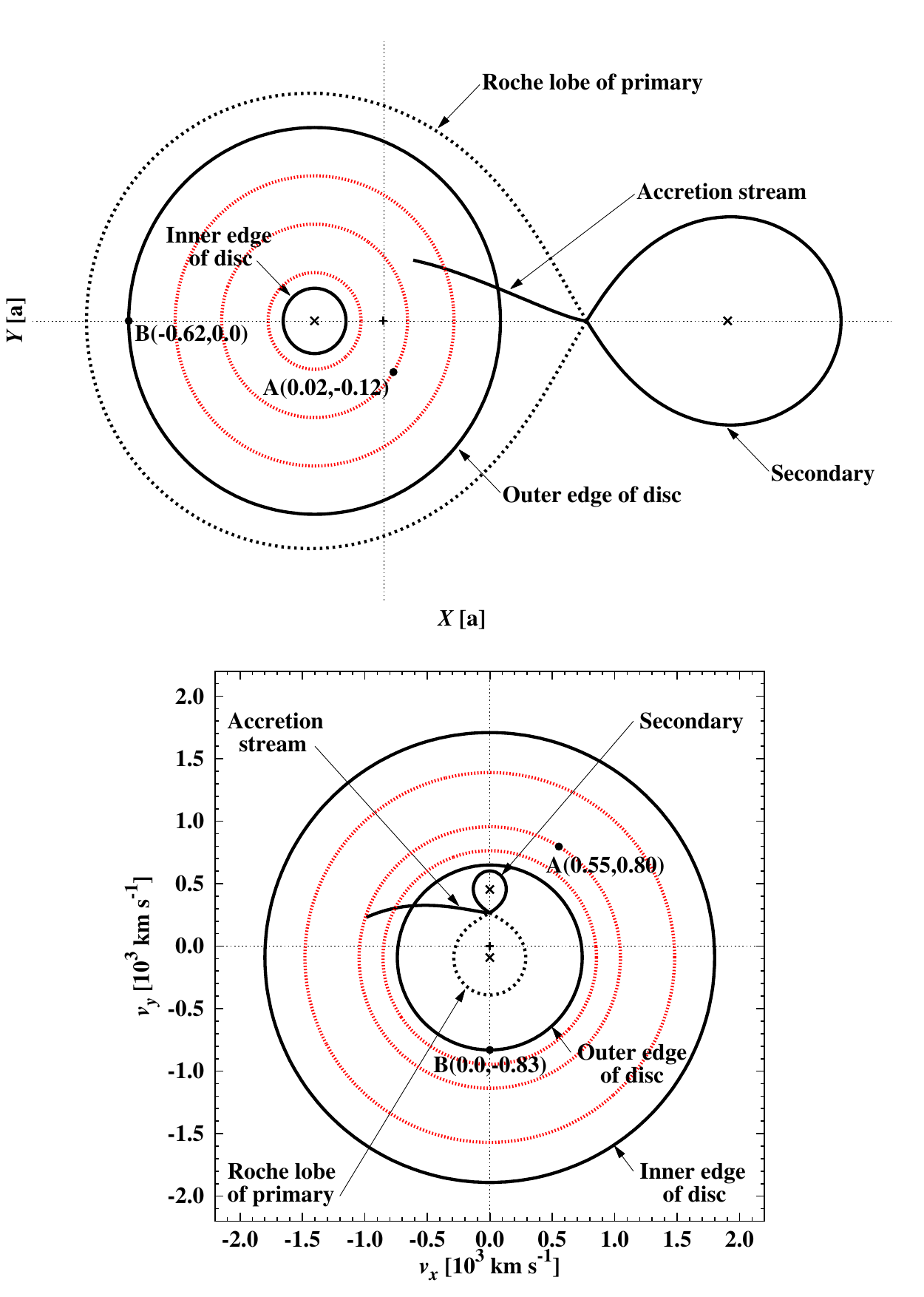}
  \caption
   {
    Standard spatial and velocity cartesian coordinates for a model CV.
    The C.O.M.~of the binary is marked with a plus (+) while that of the primary
    and secondary are marked with crosses ($\times$).
    Overlays of the Roche lobe of the primary (dashed line), the accretion
    stream (solid line) from the secondary towards the primary (up to an azimuth
    angle of $30\degr$), as well as radii at decrements of 0.25 times the outer
    disc radius are shown (red dashed circles).
    The outer disc radius is the 3:1 resonance radius while the inner disc
    radius is based on an absolute radial velocity of $1800\mbox{\,km\,s}^{-1}$.
    There are also two arbitrary reference points, A and B, shown in the disc.
   }
  \label{fig:Cartesian}
  \end{figure}
  We first review the cartesian coordinate frame that co-rotates with the binary
  system, as described by \citet{1988MNRAS.235..269M}.
  The top panel of Fig.~\ref{fig:Cartesian} shows the cartesian spatial
  coordinates (normalised by the binary separation $a$) for a model CV with an
  accretion disc.
  The assumed binary system parameters for the model CV are:
  a primary mass of $0.85M_{\sun}$, a mass ratio of $q = 0.2$, an orbital period
  of 0.05669 d ($82\mbox{\,min}$) and an inclination of $77\degr$.
  The origin is placed at the centre of mass (C.O.M.) of the binary (for a low
  $q$ system this is very close to the C.O.M.~of the primary), the $X$-axis runs
  along a line through the centres of the primary and secondary, the $Y$-axis
  runs along a line parallel to the velocity vector of the secondary and the
  $Z$-axis along a line through the C.O.M.~of the binary perpendicular to the
  orbital plane.
  In CVs where Roche-lobe overflow is considered to be responsible for the
  accretion stream and disc, these features are primarily confined to the
  orbital plane, therefore a two-dimensional layout in the orbital plane, i.e.,
  the $XY$-plane, is sufficient.
  The assumption that all motion is parallel to the orbital plane forms one of
  the postulates of Doppler tomography
  \citep{2001LNP...573....1M}.
  The binary orbital phase zero is defined as the point of mid-eclipse of the
  primary by the secondary.
  The orbital motion is counter-clockwise around the C.O.M.~of the binary.

  Every spatial point ($x,\,y$) in the system has a two-dimensional cartesian
  velocity coordinate ($v_{x},\,v_{y}$) as shown in the bottom panel of
  Fig.~\ref{fig:Cartesian}.
  Given the definition of the spatial frame it follows that, for example, the
  C.O.M.~of the primary is moving in a negative $Y$-direction ($v_{y}$ is
  negative), but with no movement in the $X$-direction ($v_{x}$ is zero).
  Similarly, the C.O.M.~of the secondary is moving in a positive $Y$-direction
  ($v_{y}$ is positive), but also with no movement in the $X$-direction
  ($v_{x}$ is zero).
  Structures such as the accretion stream and disc are more complex since they
  have a rotational velocity about the C.O.M.~of the binary and a velocity due
  to structural motion.
  The increasing velocity of the accretion stream along its ballistic trajectory
  is clearly seen in the absolute increase in its $v_{x}$ component.
  Due to the Keplerian nature of the accretion disc its velocity profile appears
  reversed, i.e., the higher velocities of the inner edge of the disc are
  farther from the origin in velocity space and conversely for the lower
  velocities of the outer edge of the disc.
  The Keplerian nature of the disc is further emphasised by the radii at
  decrements of 0.25 times the outer disc radius.
  The secondary and ballistic stream also appear to be `inside' the disc in
  velocity coordinates.

\begin{figure}
  \centering
  \includegraphics[width=8cm]{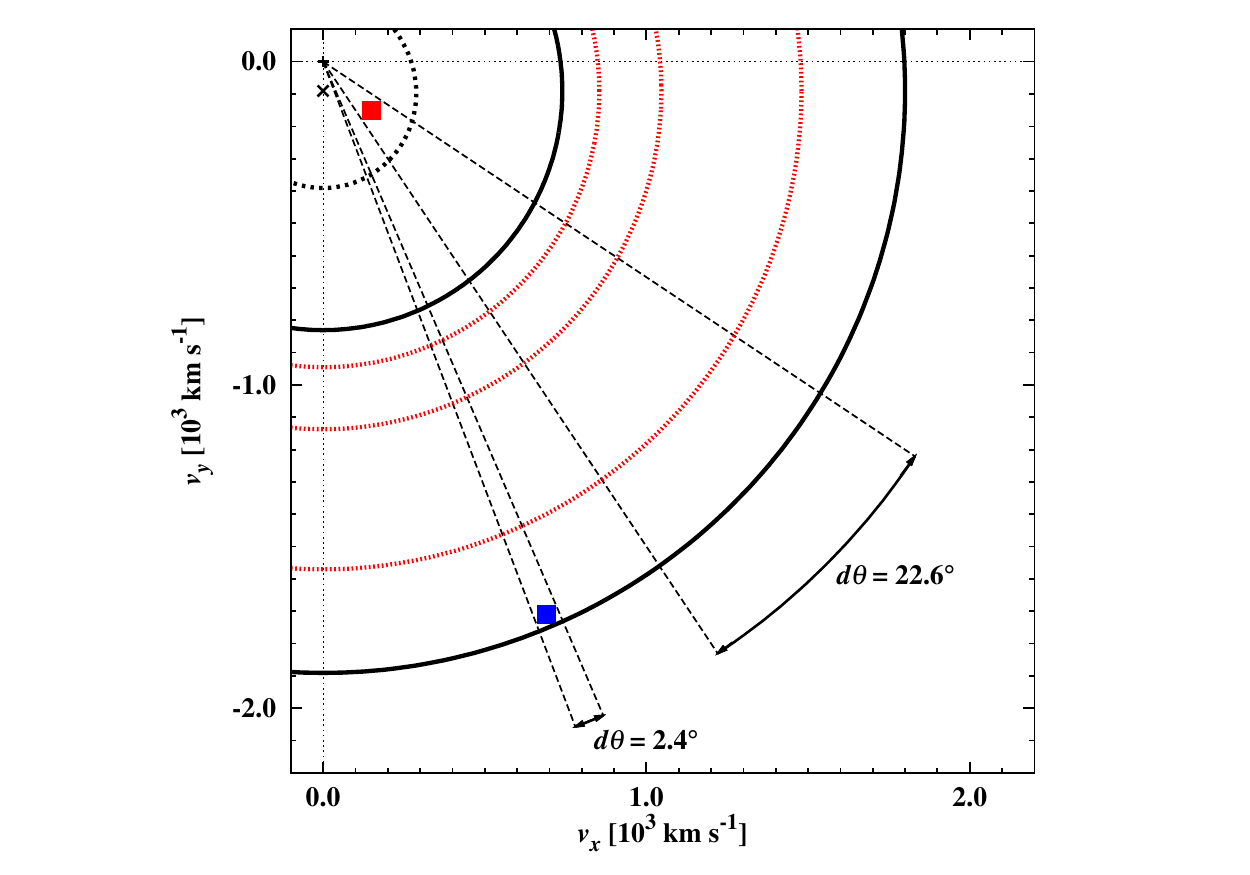}
  \caption
   {
    Velocity bins in the cartesian velocity framework (i.e., in the lower right
    quadrant of the Doppler map in Fig.~\ref{fig:Cartesian}).
    The figure shows the positions of a lower and a higher velocity pixel
    (red and blue, respectively).
   }
  \label{fig:VBinsStd}
  \end{figure}

  We note that it is not possible to invert the velocity map to produce a
  spatial map, i.e., every velocity coordinate does not have a unique spatial
  coordinate making the inversion from velocity to spatial coordinates
  impossible without additional information
  \citep[e.g.,][]{2004A&A...417.1063H}.
  In reality, orbitally phase-resolved spectroscopic observations are mapped
  directly onto the velocity map.
  Effectively each pixel, centred on a velocity coordinate ($v_{x},\,v_{y}$) in
  the velocity map, represents a velocity bin with a width ($dv_{x},\,dv_{y}$).
  In the cartesian coordinate frame of square pixels, $dv_{x} = dv_{y}$ for all
  velocity bins throughout the map.
  For example, in Fig.~\ref{fig:VBinsStd} we show a red and a blue pixel in the
  lower right quadrant of the Doppler map in Fig.~\ref{fig:Cartesian} with
  velocities ($v_{x},\,v_{y}$) = ($150,\,-150$) and ($690,\,-1710$)
  $\mbox{\,km\,s}^{-1}$, respectively.
  Both pixels have the same velocity bin widths of
  $dv_{x} = dv_{y} = 60\mbox{\,km\,s}^{-1}$ corresponding to a range in velocity
  \emph{magnitude} of $85\mbox{\,km\,s}^{-1}$ between opposite velocity
  bin corners.
  However, the range in velocity \emph{angle} ($d\theta$) is $22.6\degr$ and
  $2.4\degr$, respectively, i.e., the range in velocity angle decreases as a
  function of radial distance from the origin.
  Consequently, the lower radial velocity bins (pixels) sample a larger velocity
  vector range than the higher radial velocity bins (pixels).

  The sinusoidal radial velocity curve $v_{r}\left(\phi\right)$, traced by each
  velocity coordinate ($v_{x},\,v_{y}$), is a function of the orbital phase
  $\phi$ and centred on the mean (systemic) velocity $\gamma$ of the binary,
  which is given by:
  \begin{eqnarray}
  \label{eq:v_radial}
    v_{r}\left(\phi\right) & = & \gamma - v_{x}\cos2\pi\phi + v_{y}\sin2\pi\phi.
  \end{eqnarray}
  To simplify the discussion this can be equally given in polar coordinates as:
  \begin{eqnarray}
  \label{eq:v_radial_polar}
    v_{r}\left(\phi\right) & = & \gamma - v\cos(\theta + 2\pi\phi),
  \end{eqnarray}
  where $v$ is the velocity magnitude and $\theta$ is the velocity angle.
  The range in velocity magnitude of $85\mbox{\,km\,s}^{-1}$ between opposite
  pixel (velocity bin) corners results in four different sinusoidal amplitudes.
  The range in amplitude is the same for both the red and blue pixels.
  However, the lower velocity red pixel subtends a larger velocity angle
  ($\theta$) which corresponds to a larger range in the relative phases of the
  four sinusoids and therefore bound a larger area in the radial
  velocity-orbital phase space.
  Conversely, the blue pixel has a narrower bounding area in the radial
  velocity-orbital phase space.
  The implications of this effect is discussed in later sections.

\subsection{Velocity in polar coordinates}
\label{sec:VelPol}
  \begin{figure}
  \centering
  \includegraphics[width=8cm]{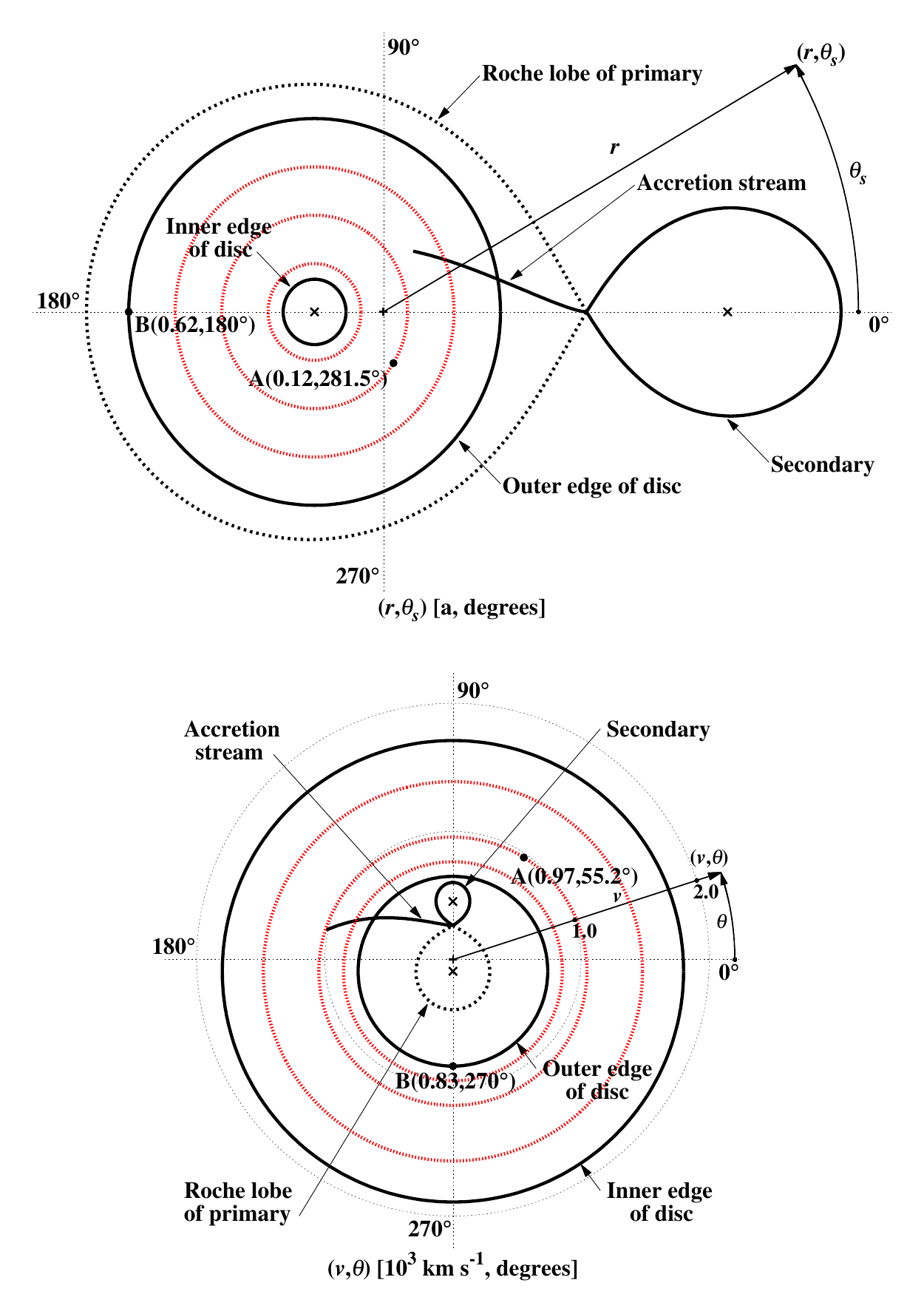}
  \caption
   {
    As for Fig.~\ref{fig:Cartesian} but in polar coordinates.
   }
  \label{fig:PolarStd}
  \end{figure}
  We continue our discussion in the polar coordinate system as this is more
  conducive to the circularly symmetric nature of Doppler tomograms and gives a
  more direct indication of velocities and directions.
  The same CV model shown in Fig.~\ref{fig:Cartesian} is shown in
  Fig.~\ref{fig:PolarStd}, but in spatial and velocity \emph{polar} coordinates.
  Spatial coordinates (normalised by the binary separation $a$) are now given as
  the radial distance $r$ from the C.O.M.~of the binary and the polar angle
  $\theta_{s}$ (not to be confused with $\theta$ in the velocity frame) measured
  in an anti-clockwise direction from the line between the C.O.M.~of the binary
  and the secondary (Fig.~\ref{fig:PolarStd} top panel).
  The corresponding velocity map in polar coordinates is shown in the bottom
  panel of Fig.~\ref{fig:PolarStd}, again with the same model CV as for
  Fig.~\ref{fig:Cartesian}.
  Velocity magnitude $v$ increases as a linear function of distance from the
  origin with a direction given by the polar angle $\theta$ in an anti-clockwise
  direction measured from the line drawn from the C.O.M.~of the binary
  horizontally to the right.
  All the CV model parameters and the arbitrary points A and B are the same as
  in the previous section.
  Therefore, for example, the secondary has a velocity magnitude of
  $400\mbox{\,km\,s}^{-1}$ with a velocity direction (angle) of $\theta=90\degr$
  (i.e., $v_{x} = 0$ and $v_{y} = 400\mbox{\,km\,s}^{-1}$ in cartesian
  coordinates).

\subsection{Velocity in inside-out polar coordinates}
\label{sec:VelIOPol}
  Having described the velocity polar coordinate frame in the previous section,
  we now investigate constructing an inside-out velocity polar coordinate frame
  with zero velocity on the outer circumference and the maximum velocities
  around the centre of the coordinate frame.
  The velocity magnitude $v$ now increases as a linear function of distance from
  the zero velocity outer circumference towards the origin with a direction
  given by the polar angle $\theta$.
  In order to make the inside-out framework complementary to the standard
  framework, and to facilitate a direct comparison between the frameworks, the
  default origin of the inside-out framework is set to the maximum radial
  velocity used to construct the standard framework.
  Effectively this is the maximum radial velocity extracted from the
  phase-resolved spectra of a source emission (or absorption) line, i.e., the
  `edge' of the data projected onto the framework to construct the tomogram.

  \begin{figure}
  \centering
  \includegraphics[width=8cm]{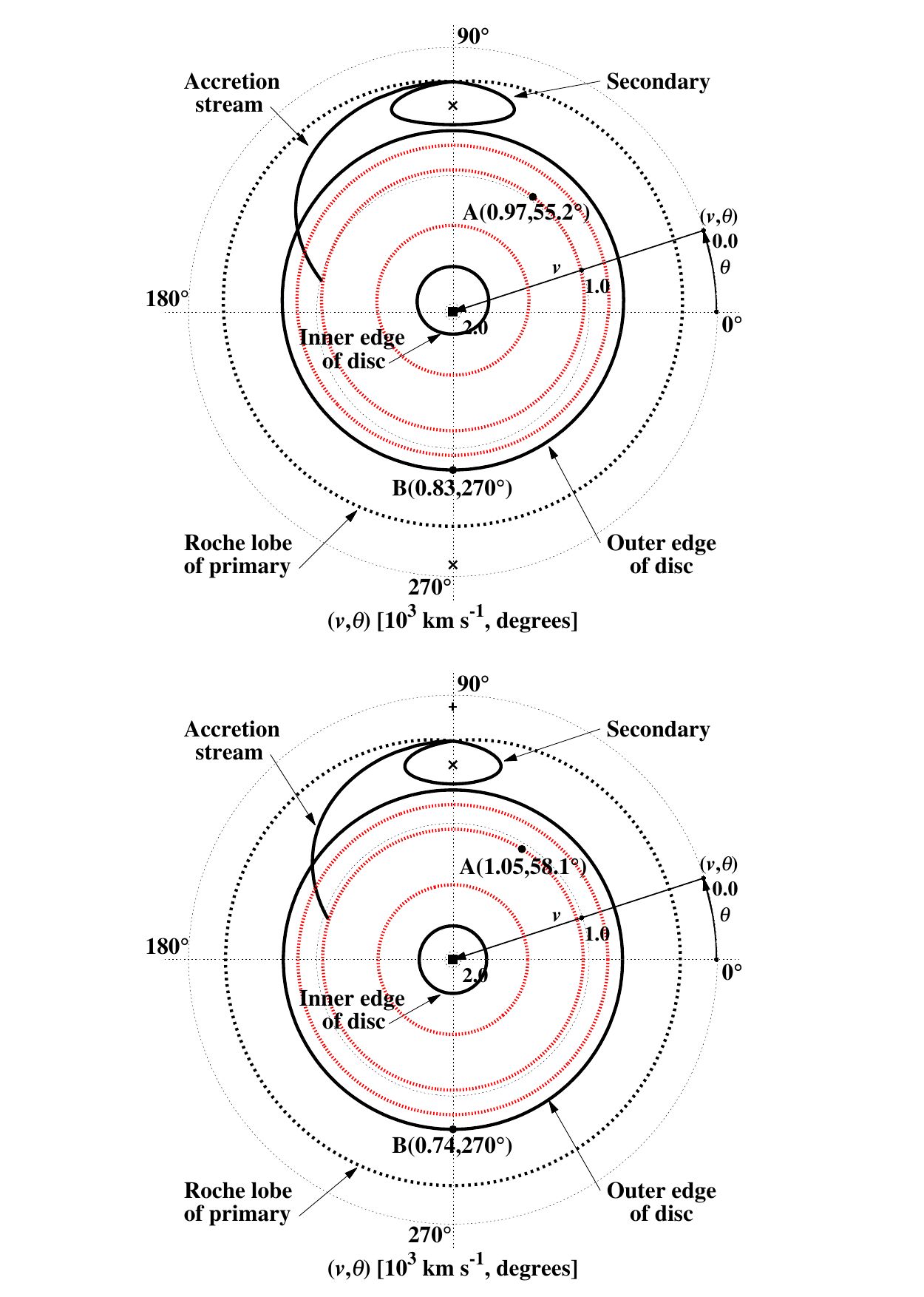}
  \caption
   {
    Inside-out velocity polar coordinates.
    The top and bottom panels show the model CV centred on the anti-C.O.M.~of
    the binary and the anti-C.O.M.~of the primary, respectively.
    See text for full description.
   }
  \label{fig:PolarIo}
  \end{figure}
  The inside-out velocity map of the model CV presented in the previous figures
  is shown in the top panel of Fig.~\ref{fig:PolarIo}.
  Essentially the zero velocity origin (corresponding to the C.O.M.~of the
  binary) is now transposed to the outer circumference and the maximum outer
  velocities to the central origin, giving the new inside-out velocity
  framework.
  The central origin, however, represents a discontinuity and the inner 4
  velocity bins (pixels) around it are therefore ignored.
  We would like to point out that the spatial map (Fig.~\ref{fig:PolarStd} top
  panel) remains unchanged.
  As one can see (Fig.~\ref{fig:PolarIo} top panel) the inner and outer edges of
  the accretion disc are now correctly orientated.
  Also the ballistic stream is now on the outside of the disc and `curves'
  inwards as it accelerates towards the disc and primary.
  The Roche lobe of the secondary is also now on the outside of the disc albeit
  upside down.
  It is upside down because it is orbiting as a solid body with the outside
  moving faster than the inside, as opposed to the Keplerian velocities of an
  accretion disc and the ballistic velocities of the accretion stream.
  The Roche lobe of the primary is now the outer bounded circular ring area,
  i.e., between zero velocity and the dashed line, that also contains the
  C.O.M.~of the primary (indicated by a $\times$).

  The Keplerian velocity radial profile of the disc is still visibly apparent as
  can be seen by the equidistant disc radii contours (Fig.~\ref{fig:PolarIo} top
  panel).
  We explored replacing the linear velocity axes with a Keplerian velocity
  ($\propto1/\sqrt{r}$) profile in an attempt to produce a more equidistant
  velocity contour profile.
  However, a correct calculation of the Keplerian profile would require
  knowledge of the primary mass of the actual system under investigation.
  This is rarely known and would introduce a new parameter that would no longer
  give a map that is simply a direct projection of the observations.

  We note that the inner edge of the accretion disc is not centred on the origin
  (Fig.~\ref{fig:PolarIo} top panel).
  This is because the disc is centred on the C.O.M.~of the primary, whereas the
  origin represents the \emph{anti}-C.O.M.~of the binary, i.e., the C.O.M.~of
  the binary is effectively the outer circumference of zero velocity with its
  anti-C.O.M.~at the origin.
  Hence the Keplerian velocities of the disc will be offset by an amount
  corresponding to the orbital motion of the primary.
  This offset is also present in standard velocity maps but is more apparent in
  the inside-out framework.
  The bottom panel of Fig.~\ref{fig:PolarIo} shows the same model but with the
  orbital motion of the primary subtracted from the model velocity profile.
  This effectively places the primary on the zero velocity outer circumference
  and its anti-C.O.M.~at the origin.
  The effects of centring Doppler tomograms on the C.O.M.~of the binary or of
  the primary are investigated in the next section.

  \begin{figure}
  \centering
  \includegraphics[width=8cm]{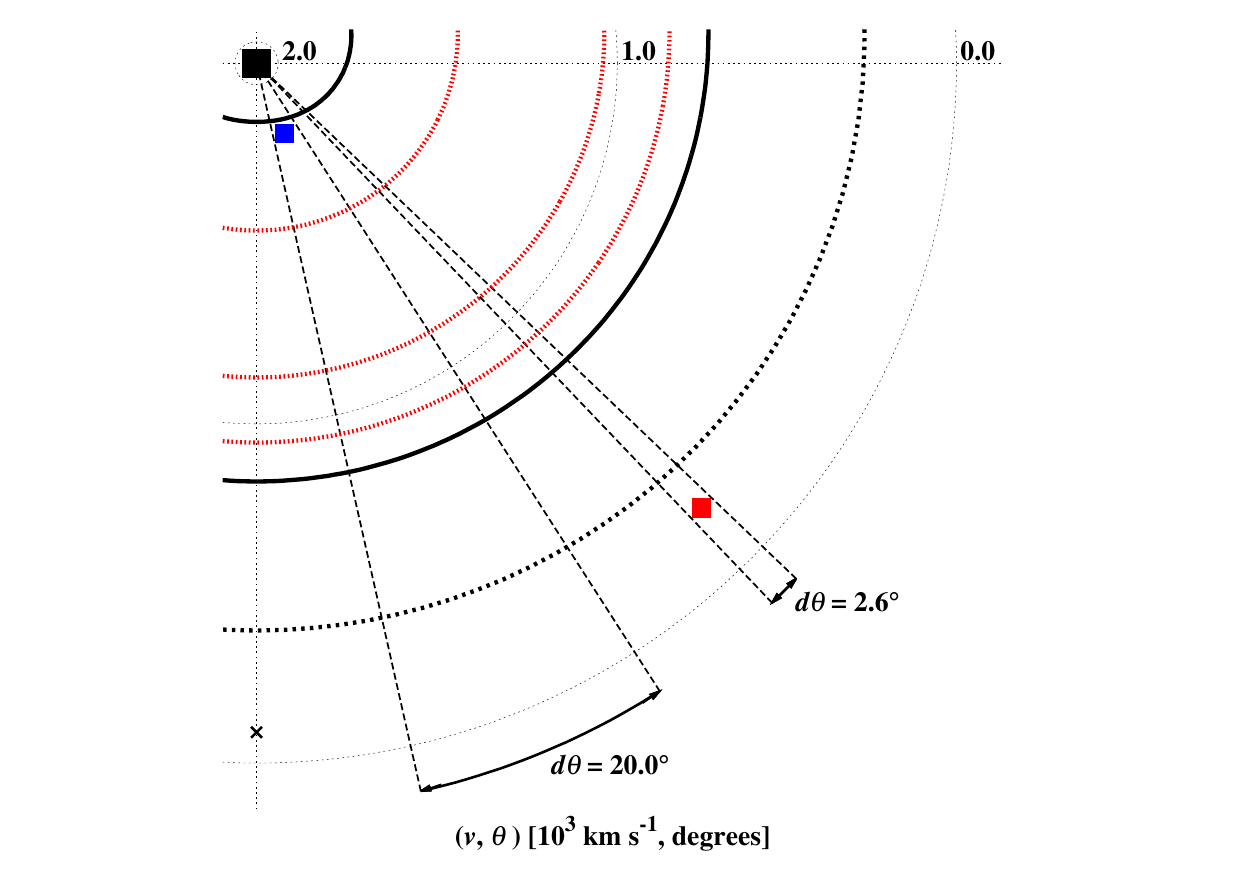}
  \caption
   {
    Velocity bins in the inside-out polar velocity framework (i.e., in the lower
    right quadrant of the Doppler map in the top panel of
    Fig.~\ref{fig:PolarIo}).
    The figure shows the positions of a lower and a higher velocity pixel
    (red and blue, respectively).
    These pixels are the inside-out polar equivalent of the cartesian pixels
    shown in Fig.~\ref{fig:VBinsStd}.
   }
  \label{fig:VBinsIo}
  \end{figure}
  As described in Sect.~\ref{sec:VelCart} each velocity bin (square pixel) has
  the same velocity \emph{magnitude range}; however, the velocity \emph{vector
  range} decreases as a function of radial distance from the central origin.
  To demonstrate this for the inside-out framework we show in
  Fig.~\ref{fig:VBinsIo} the equivalent red and blue pixels shown in
  Fig.~\ref{fig:VBinsStd}.
  In the inside-out framework the higher velocity pixels are closer to the
  origin and therefore sample a larger velocity vector range than in the
  standard framework and conversely for the lower velocity pixels.
  This can be seen clearly by comparing Figs \ref{fig:VBinsStd} and
  \ref{fig:VBinsIo}.
  The implications of this effect are discussed in the following sections.

\section{Doppler tomography: standard and inside-out projections}
\label{sec:DopTom}

  For real CVs, the orbital and structural velocities can be observed in the
  Doppler-shifted variations in the multi-component emission (and absorption)
  lines of their optical spectra.
  The orbitally phase-resolved spectroscopy of individual spectral lines can be
  projected onto a predefined velocity framework to produce a Doppler tomogram.
  The interpretation of such tomograms would then proceed by comparison with
  overlays of model velocity profiles of the likes described in the previous
  section
  \citep[e.g.,][]{1994MNRAS.266..137M,1998A&A...332..984W}.
  Traditionally, Doppler tomograms are produced by a projection onto a cartesian
  coordinate frame, but given the circularly symmetric velocity profile of
  binary systems we elected to use a polar coordinate frame for our standard and
  inside-out tomograms as discussed in earlier sections.

  We have modified the fast maximum entropy Doppler tomography code presented by
  \citet{1998astro.ph..6141S}
  to incorporate the inside-out velocity framework\footnote{Our code may be
  obtained upon request.}.
  With the modified code the observed Doppler-shifted spectra are projected
  directly onto an inside-out velocity coordinate frame.
  All the standard and inside-out Doppler tomograms presented hereafter have
  been created using the unmodified and modified code, respectively.

\subsection{Simulated system: accretion disc, bright spot and secondary}
\label{sec:SynCV}
  \begin{figure*}
  \centering
  \includegraphics[width=17cm]{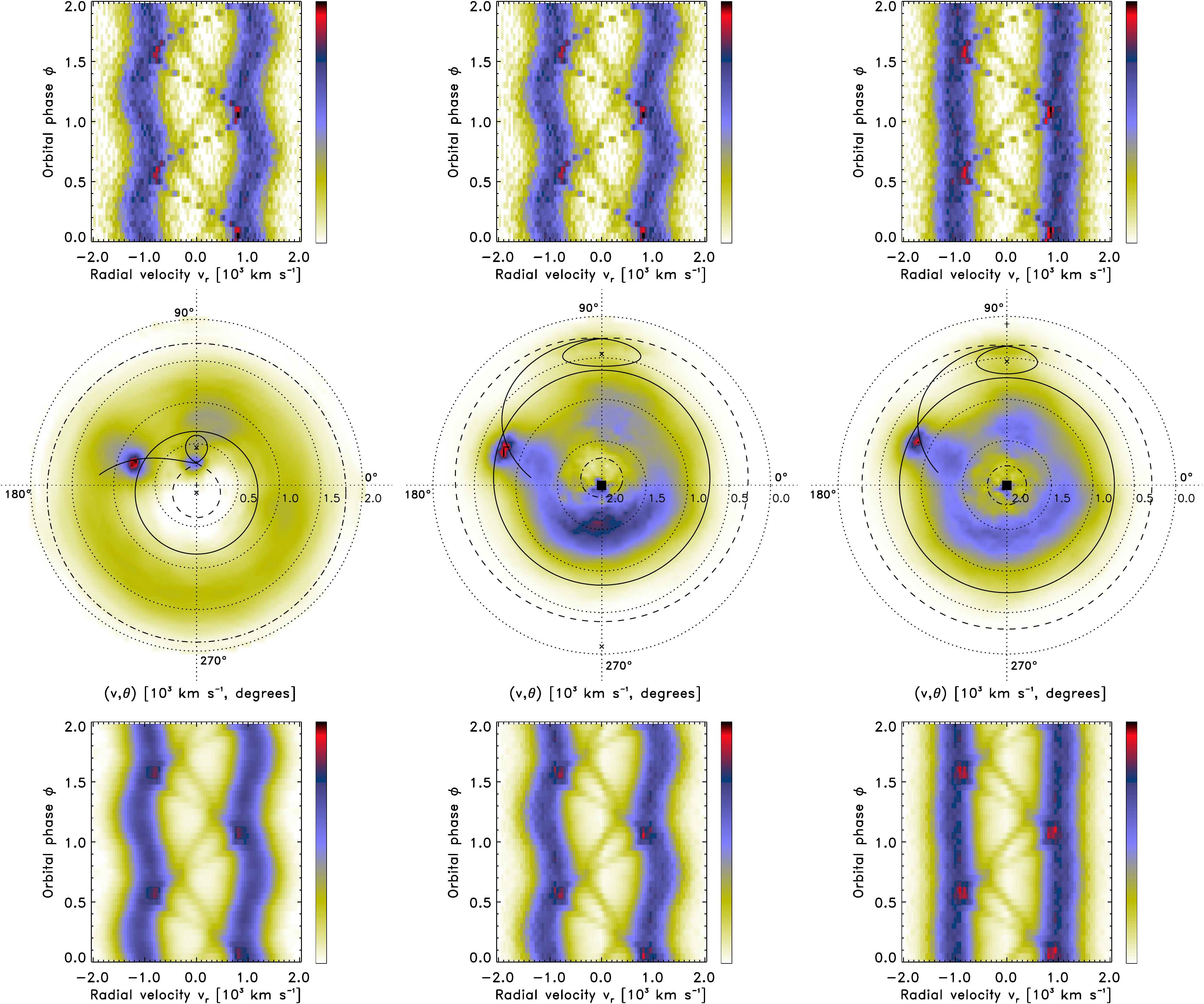}
  \caption
   {
    Doppler tomography of synthetic spectra.
    The standard tomogram and two inside-out tomograms (with the anti-C.O.M.~of
    the binary and of the primary at the origin, respectively) are shown from
    left to right in the middle panels for comparison (see text for full
    description).
    The top and bottom panels show the input and reconstructed trailed spectra,
    respectively, for the corresponding tomogram.
   }
  \label{fig:SynCV}
  \end{figure*}

  The results of applying standard and inside-out Doppler tomography to a
  simulated data set are shown in Fig.~\ref{fig:SynCV}.
  A single line emission spectrum was constructed to simulate a disc-accreting
  system with an orbital period of 0.05669 d ($82\mbox{\,min}$), a mass ratio of
  $q = 0.2$, a primary mass of $0.85M_{\sun}$ and an inclination of $77\degr$.
  The emission line was made of four components: two relatively narrow Gaussian
  profiles to represent emission from the accretion disc's bright spot and the
  irradiated face of the secondary star, a broad double peaked profile to
  represent emission from an accretion disc and some artificial noise.
  Appropriate phase-dependent Doppler shifts were then calculated for each of
  the components before summing to produce the trailed spectra shown in the top
  left panel of Fig.~\ref{fig:SynCV}.
  The effect of inclination was also taken into account.

  The binary system parameters listed above were used to calculate the model
  velocity profile overlay shown in all the tomograms.
  Included in the overlay are the Roche lobes of the primary (dashed line), the
  secondary (solid line) and a single particle ballistic trajectory (solid line)
  from the $L_{1}$ point up to $40\degr$ in azimuth around the primary, as well
  as the 3:1 resonance radius as the outer edge of the accretion disc (large
  solid line circle).
  Also shown is an inner Keplerian radius (dot-dashed line) based on an absolute
  radial velocity of $1800\mbox{\,km\,s}^{-1}$.
  This represents the maximum absolute radial velocity seen in the emission line
  variations.

\subsubsection{Standard Doppler tomogram}
  The middle left panel of Fig.~\ref{fig:SynCV} shows the standard Doppler
  tomogram.
  The ring-like feature corresponds to emission from the disc with additional
  emission regions at the locations of the bright spot and secondary star.
  The lower left panel is the reconstructed trailed spectra from the Doppler
  code.

\subsubsection{Inside-out Doppler tomogram}
\paragraph{Anti-C.O.M.~of binary at origin:}
  The same trailed spectra (top middle panel) were used to produce the
  inside-out Doppler tomogram (middle panel).
  As expected from our modelling in previous sections, emission from the
  secondary star is now located on the outside of the disc and the bright spot
  on the disc's outer edge.
  Furthermore the disc has the `correct' orientation with a radial profile that
  has changed from a more even distribution in the standard tomogram, to a more
  inner disc concentration.
  This can be understood in terms of our discussion in Sect.~\ref{sec:VelSpace},
  i.e., in the standard tomogram the higher velocity disc emission is
  distributed over more pixels covering a larger surface area.
  In the inside-out tomogram the higher velocity disc emission is concentrated
  into fewer pixels closer to the origin and hence appears brighter.
  This is perhaps more consistent with what one might expect from disc emission,
  i.e., the emitting volume in the inner edges are much smaller that the outer
  edges of the disc.
  However, the lower half of the disc appears to contain more emission at higher
  velocities than the upper half.
  This asymmetric artefact arises because the \emph{Keplerian motion} of the
  accretion disc orbits the primary which in turn orbits the C.O.M.~of the
  binary, i.e., the disc is projected off-centre.
  This asymmetry is also present in the standard tomograms but is enhanced in
  the inside-out tomograms due to the brightening of the high-velocity pixels
  around the origin.

\paragraph{Anti-C.O.M.~of primary at origin:}
  We next subtracted the orbital velocity of the primary
  $K_{1}\sim91\mbox{\,km\,s}^{-1}$
  \citep[e.g.,][]{1987MNRAS.225..551M}
  from the phase-resolved spectra (top right panel).
  As expected, the double peak emission from the disc has been `straightened' in
  the trailed spectra, i.e., with the orbital velocity of the primary removed,
  emission from a given radius in the disc shows a constant radial velocity in
  the trailed spectra.
  In the recalculated inside-out tomogram (middle right panel) the C.O.M.~of the
  primary now effectively becomes the outer circumference of zero velocity with
  its anti-C.O.M.~at the origin.
  Consequently, the disc is now centred on the origin and the asymmetry seen in
  the disc emission has disappeared as expected.

\subsection{WZ Sge in quiescence: accretion disc and bright spot}
\label{sec:WZSge}
  \begin{figure*}
  \centering
  \includegraphics[width=17cm]{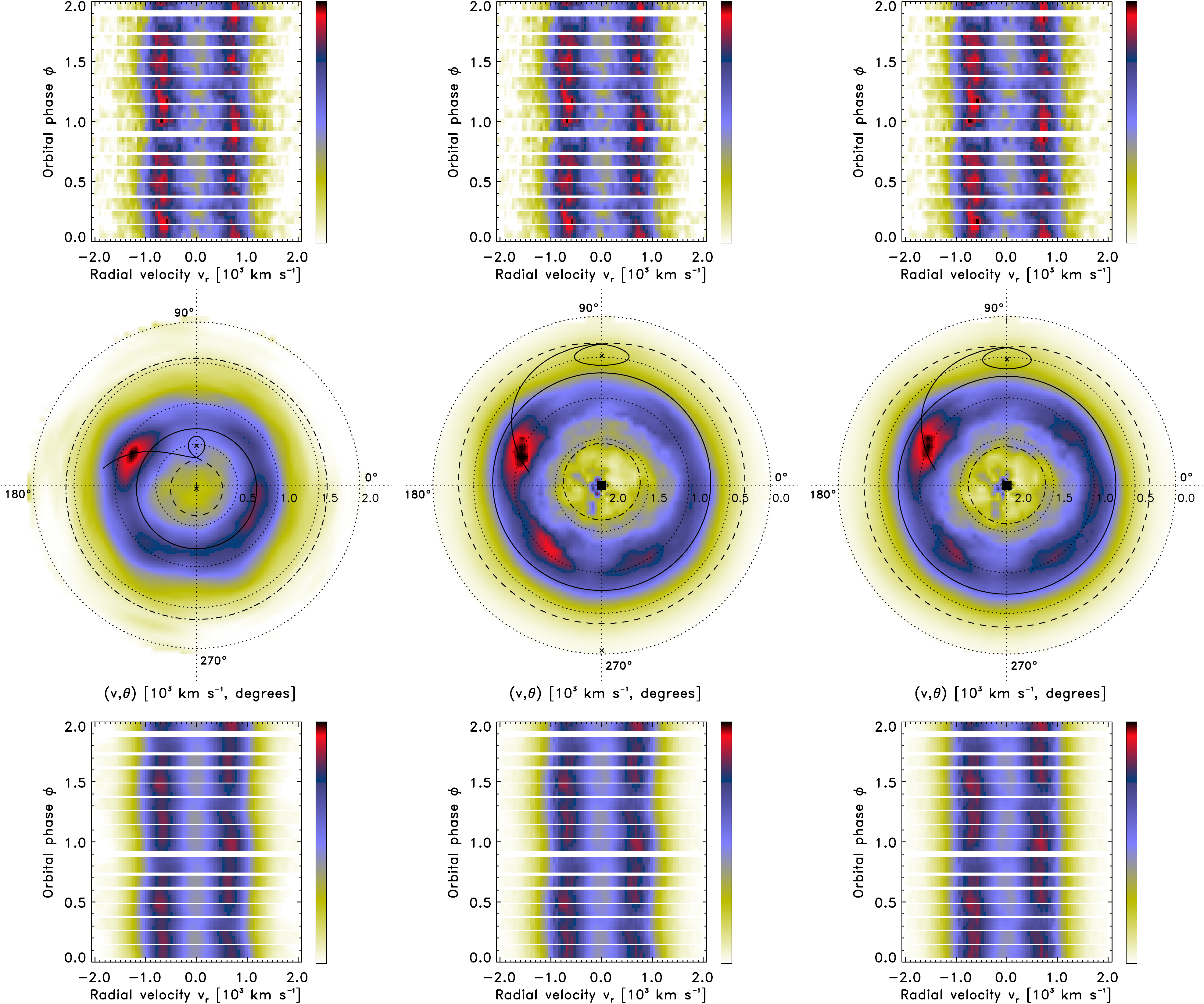}
  \caption
   {
    Doppler tomography of WZ Sge in quiescence.
    Same layout as Fig.~\ref{fig:SynCV}.
   }
  \label{fig:WZSge}
  \end{figure*}

  WZ Sge is a prototypical member of the CV subclass referred to as WZ Sge-type
  dwarf novae
  \citep{1979MNRAS.189P..41B}.
  Doppler tomography of WZ Sge
  \citep[e.g.,][]{1998MNRAS.299..768S}
  has revealed a prominent accretion disc and bright spot where the ballistic
  accretion stream impacts the outer edge of the disc.
  This makes it an ideal test case for inside-out Doppler tomography.

  The standard and inside-out Doppler tomography results for WZ Sge in
  quiescence are shown in Fig.~\ref{fig:WZSge} (similar layout to
  Fig.~\ref{fig:SynCV}).
  The tomography is based on the \halpha emission line from spectroscopic
  observations included with the fast maximum entropy Doppler tomography code
  \citep{1998astro.ph..6141S}.
  The model velocity profile overlay shown in all the tomograms was calculated
  using the orbital period of $0.056687846$ d ($82\mbox{\,min}$)
  determined by
  \citet{1998PASP..110..403P},
  the inclination of $77\degr$ obtained by
  \citet{1998MNRAS.299..768S}
  and the mass parameters preferred by
  \citet{2007ApJ...667..442S},
  i.e., a primary mass of $0.85M_{\sun}$ and a mass ratio $q = 0.09$.
  The overlay includes the Roche lobes of the primary (dashed line), the
  secondary (solid line) and a single particle ballistic trajectory (solid line)
  from the $L_{1}$ point up to $45\degr$ in azimuth around the primary, as well
  as the 3:1 resonance radius as the outer edge of the accretion disc (large
  solid line circle).
  An inner Keplerian radius (dot-dashed line) based on an absolute radial
  velocity of $1600\mbox{\,km\,s}^{-1}$ which represents the maximum absolute
  radial velocity seen in the emission line variations, is also shown.

  \subsubsection{Standard Doppler tomogram}
  The most distinct features in the standard tomogram shown in the middle left
  panel of Fig.~\ref{fig:WZSge} are the relatively bright accretion disc and the
  bright spot of the impact region between the ballistic stream and disc.
  The disc appears brighter around the assumed outer edge (velocities lower than
  $1000\mbox{\,km\,s}^{-1}$) while it becomes more diffused towards the assumed
  inner edge (velocities greater than $1000\mbox{\,km\,s}^{-1}$).
  The extended nature of the bright spot (stream impact region) in velocity
  space is considered to be the result of the mixing of stream and disc material
  with different velocities
  \citep{1990ApJ...364..637M}.
  The model stream lines up well with the brightest part of the impact region
  located at ($820\mbox{\,km\,s}^{-1},\,150\degr$).
  There is only a slight enhancement in the emission at the expected velocity of
  the secondary ($400\mbox{\,km\,s}^{-1},\,90\degr$).

\subsubsection{Inside-out Doppler tomogram}
\paragraph{Anti-C.O.M.~of binary at origin:}
  In the inside-out tomogram (middle panel) the disc is now correctly orientated
  with brighter disc emission occurring at generally higher velocities
  ($>1000\mbox{\,km\,s}^{-1}$).
  Due to the effect of the off-centre projection of the disc there is a slight
  enhancement in the brightness of the disc in the lower half.
  Although given WZ Sge's low mass ratio this effect is fairly minimal.
  Similar to the standard tomogram the bright spot (stream impact region) also
  dominates the brightness scale of the inside-out tomogram.
  However, it is now located towards the disc's outer edge in velocity space and
  has acquired an extended enhancement in the lower left quadrant covering
  ($1000\mbox{\,km\,s}^{-1},\,180-250\degr$).
  The brightest part of the impact region it is now located at
  ($1050\mbox{\,km\,s}^{-1},\,165\degr$), a slightly different velocity position
  than in the standard tomogram.
  This shift is caused by the redistribution of the relative contrast levels
  throughout the tomogram.
  The model stream, however, still lines up well with the brightest part of the
  impact region.
  The secondary is not discernible within the low-velocity emission covering
  ($0-700\mbox{\,km\,s}^{-1},\,0-360\degr$).
  This is expected since it was barely discernible in the standard tomogram and
  is projected into more pixels in the inside-out tomogram.
  However, by subtracting the axisymmetric average of the emission at each
  radius around the origin (not shown), there is a patch of diffuse emission at
  the expected location of the secondary.
  Similar to the simulated system in the previous section, the high-velocity
  ($2000\mbox{\,km\,s}^{-1}$) noise present at the edge of the spectra is
  enhanced by the brightening of the high-velocity pixels around the origin.

\paragraph{Anti-C.O.M.~of primary at origin:}
  The velocity of the primary $K_{1}\sim47\mbox{\,km\,s}^{-1}$ with a phase zero
  offset $0.12$
  \citep{2007ApJ...667..442S}
  was subtracted to obtain the phase-resolved spectra in the top right panel of
  Fig.~\ref{fig:WZSge}.
  Given the relative low velocity of the primary that was subtracted, the
  `straightening' of the double peak emission from the disc in the trailed
  spectra is less pronounced compared to the case of the simulated system in the
  previous section.
  In the recalculated inside-out tomogram (middle right panel) the anti-C.O.M.~of
  the primary is now at the origin.
  With the effect of the off-centre projection of the disc removed a more
  uniform circularly symmetric appearance of the disc is recovered.
  The extended enhancement of the bright spot in the lower left quadrant
  covering ($1000\mbox{\,km\,s}^{-1},\,180-250\degr$) is also less pronounced.

\subsection{IP Peg in outburst: spiral shocks}
\label{sec:IPPeg}
  \begin{figure*}
  \centering
  \includegraphics[width=17cm]{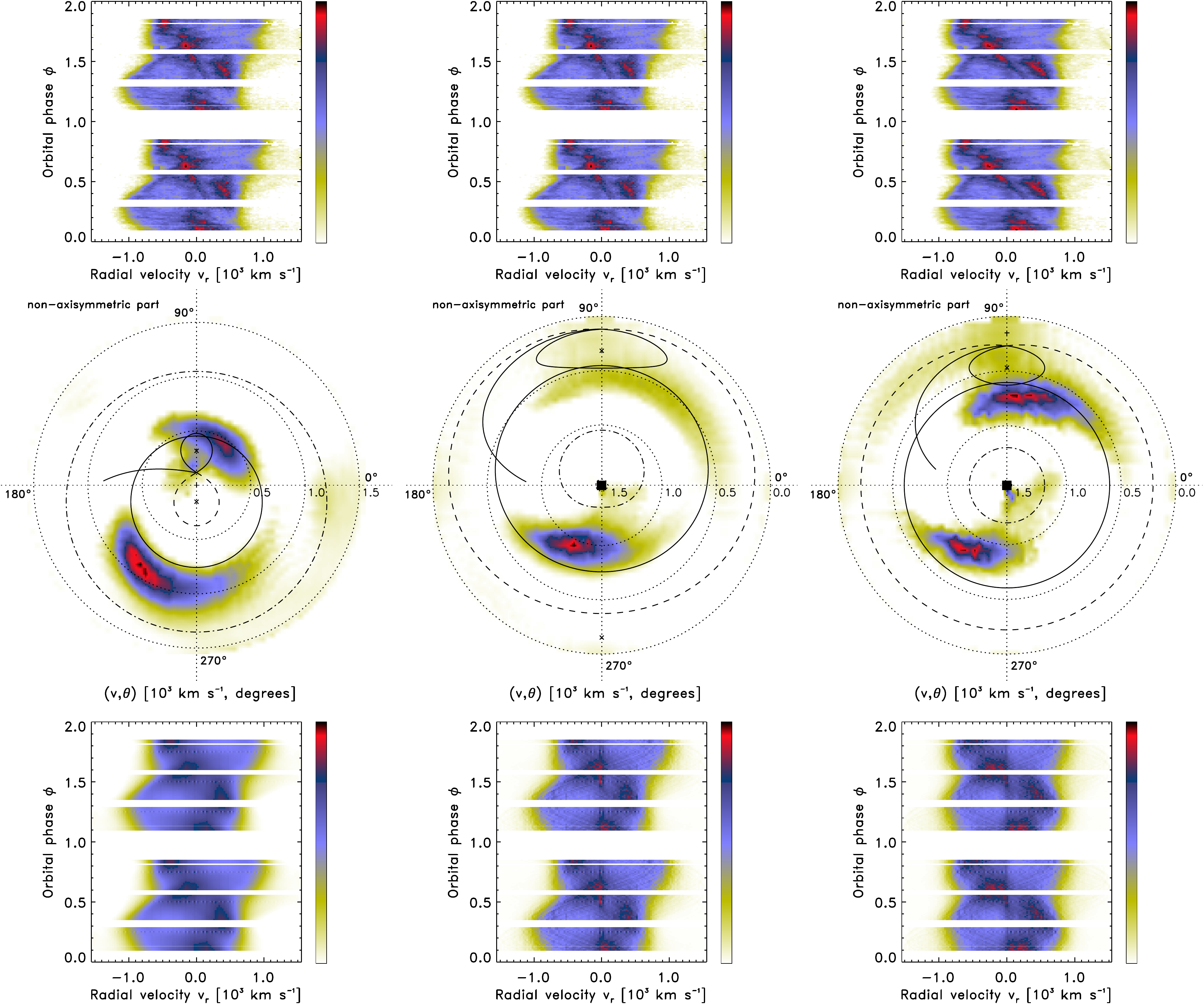}
  \caption
   {
    Doppler tomography of IP Peg in outburst.
    Same layout as Fig.~\ref{fig:WZSge} and Fig.~\ref{fig:SynCV}.
    Only the non-axisymmetric emission is shown in the tomograms.
   }
  \label{fig:IPPeg}
  \end{figure*}

  IP Peg is an eclipsing member of the dwarf nova subclass of CVs.
  Outbursts lasting $10-12$ days occur every $70-100$ days during which it
  brightens by $2-3$ magnitudes.
  The spiral shocks observed in the accretion disc of IP Peg during an outburst
  \citep{1997MNRAS.290L..28S}
  favour the disc instability model
  \citep{1974PASJ...26..429O}
  as the cause of the recurrent outbursts.
  Since the disc expands during the outburst the outer disc will experience an
  increased gravitational attraction from the secondary.
  This causes perturbations in the circular Keplerian orbits of the disc
  material in the outer disc, leading to the formation of spiral arms in the
  disc.
  The spiral structure in the accretion disc of IP Peg makes it an ideal test
  case for inside-out Doppler tomography.

  Standard and inside-out Doppler tomography results based on the \heII emission
  line from the 1996 November spectroscopic observations taken during outburst
  maximum, are shown in Fig.~\ref{fig:IPPeg} (similar layout to
  Fig.~\ref{fig:WZSge} and Fig.~\ref{fig:SynCV}).
  These observations were presented first by
  \citet{1999MNRAS.306..348H}
  and the reader is referred to that paper for a detailed description of the
  data set and reduction procedures.
  All the tomograms in Fig.~\ref{fig:IPPeg} show only the non-axisymmetric
  emission.
  The binary system parameters as calculated by
  \citet{2010MNRAS.402.1824C},
  i.e., an orbital period of $0.1582061029$ d ($228\mbox{\,min}$), an
  inclination of $83.8\degr$, a primary mass of $1.16M_{\sun}$ and a mass ratio
  $q = 0.48$ were used to calculate the model velocity profile overlay shown in
  all the tomograms.
  Included in the overlay are the Roche lobes of the primary (dashed line), the
  secondary (solid line) and a single particle ballistic trajectory (solid line)
  from the $L_{1}$ point up to $45\degr$ in azimuth around the primary, as well
  as the tidal radius as the outer edge of the accretion disc (large solid line
  circle).
  Also indicated is an inner Keplerian radius (dot-dashed line) based on an
  absolute radial velocity of $1200\mbox{\,km\,s}^{-1}$, representing the
  maximum absolute radial velocity seen in the emission line variations.

\subsubsection{Standard Doppler tomogram}
  In the standard tomogram shown in the middle left panel of
  Fig.~\ref{fig:IPPeg} the emission from the leading side of the irradiated
  secondary is seen as a bright spot at ($200\mbox{\,km\,s}^{-1},\,95\degr$)
  which appears to be `inside' the disc.
  There is also a low-velocity component visible at
  ($0-300\mbox{\,km\,s}^{-1},\,140-250\degr$).
  This low-velocity component has been seen in other dwarf novae during
  outburst, but its origin is not understood
  \citep{1996MNRAS.281..626S}.
  The origin of the high-velocity component seen as a diffuse patch of emission
  at ($1000-1500\mbox{\,km\,s}^{-1},\,0\degr$) is also uncertain.
  The tomogram is dominated by the emission from the shocks in the accretion
  disc that creates an extended two-armed spiral structure.
  The two spiral shocks have a clear asymmetry in both brightness and velocity.
  The observed velocity asymmetry of the material in the shocks suggests a
  highly non-Keplerian flow in the disc since circular rings of emission are
  expected from a purely Keplerian disc.
  The shocks appear to be spiralling `outwards' from lower to higher velocities.
  The upper spiral shock extends to velocities lower than our velocity profile
  overlay of the outer disc radius.
  This supports the suggestion by
  \citet{1997MNRAS.290L..28S}
  that the disc possibly expanded beyond its tidal radius early on in the
  outburst as a result of it being triggered by a disc instability.

\subsubsection{Inside-out Doppler tomogram}
\paragraph{Anti-C.O.M.~of binary at origin:}
  IP Peg's mass ratio (0.48) is fairly high. Consequently the C.O.M.~of the
  primary is significantly offset from the C.O.M.~of the binary system.
  This can be clearly seen in the offset of the model overlay w.r.t.~the origin
  in the inside-out tomogram (middle panel).
  As expected the secondary is seen as an extended diffuse patch of emission
  `outside' the disc.
  However, the low-velocity component visible at
  ($0-300\mbox{\,km\,s}^{-1},\,140-250\degr$) in the standard tomogram is no
  longer discernible.
  The high-velocity patch of emission at
  ($1000-1500\mbox{\,km\,s}^{-1},\,0\degr$) has become more enhanced and compact
  given its new projection closer to the origin.
  The two-armed spiral structure associated with the shocks in the disc is still
  clearly present in the inside-out tomogram.
  However, the asymmetry in the brightness of the shocks caused by the
  off-centre projection of the disc is more pronounced than in the standard
  tomogram.
  The upper shock is more extended and diffuse while the lower shock is more
  compact and brighter.

\paragraph{Anti-C.O.M.~of primary at origin:}
  The phase-resolved spectra in the top right panel of Fig.~\ref{fig:IPPeg} were
  obtained by subtracting the orbital velocity of the primary
  $K_{1}\sim152\mbox{\,km\,s}^{-1}$
  \citep[average from][]{2010MNRAS.402.1824C}.
  Given the complex structure of the trailed spectra
  \citep{1999MNRAS.306..348H}
  the `straightening' of the double peak emission from the disc is primarily
  seen in the blue peak.
  These `straightened' spectra were used as input to produce the tomogram in the
  middle right panel.
  The model overlay clearly displays the removal of the C.O.M.~offset.
  The secondary has become more compact and brighter.
  This is because the position of the secondary has moved significantly closer
  to the origin leading to its enhancement.
  Similarly the patch of high-velocity emission at
  ($1000-1500\mbox{\,km\,s}^{-1},\,0\degr$) has become slightly more enhanced.
  Although the origin of this feature is unknown it is associated with the low
  brightness emission seen in the input trailed spectra at high radial velocity
  ($>1000\mbox{\,km\,s}^{-1}$) between phases $0.4$ and $0.8$.
  In addition an extended low-velocity component is now visible at
  ($0-250\mbox{\,km\,s}^{-1},\,110-150\degr$).
  This low-velocity component is part of the low-velocity component of unknown
  origin
  \citep{1996MNRAS.281..626S}
  covering ($0-300\mbox{\,km\,s}^{-1},\,140-250\degr$) in the standard tomogram.
  The asymmetry between the brightness of the two shocks caused by the
  off-centre projection of the disc has also been removed.
  The spiral nature of the shocks is clearly evident and is more intuitive to
  interpret in respect that it shows how they begin at the outer edges of the
  disc and curve inwards towards the primary as they increase in velocity.

\section{Summary}
\label{sec:Summary}

  We have investigated the use of an inside-out velocity coordinate frame for
  Doppler tomograms of non-magnetic cataclysmic variables.
  In the inside-out framework zero velocity is on the outer circumference and
  the maximum velocities are around the centre of the coordinate frame.
  We modified the fast maximum entropy Doppler tomography code presented by
  \citet{1998astro.ph..6141S}
  to incorporate the inside-out framework which allowed us to construct
  inside-out tomograms independently of standard tomograms by directly
  projecting the observed spectra onto the inside-out velocity coordinate frame.

  We applied our new inside-out velocity projection to real data of the dwarf
  nova system WZ Sge during a normal faint state.
  The accretion disc has the correct orientation with the bright spot located
  towards the disc's outer edge (in velocity space).
  In addition there is a redistribution of disc brightness, i.e., in the
  standard tomogram the disc appears to brighten for velocities lower than
  $1000\mbox{\,km\,s}^{-1}$ whereas with the inside-out velocity projection the
  disc brightens for velocities higher than $1000\mbox{\,km\,s}^{-1}$.
  The difference arises because the line emission from the higher velocities
  emanating from the inner regions of the disc are spread out over a larger area
  with more pixels in the standard tomogram, whereas the higher velocity
  emission is concentrated in a smaller area with fewer pixels in the
  inside-out tomogram as is more in line with a real disc brightness profile.
  Given the relative low mass ratio ($q\sim0.1$) of WZ Sge this real data set
  only marginally highlighted the vertical asymmetry in the brightness
  distribution in tomograms with an off-centre projection of the accretion disc,
  i.e., with the C.O.M.~or anti-C.O.M.~of the binary at the origin.
  The asymmetry is somewhat enhanced in the inside-out velocity framework
  (anti-C.O.M.~of the binary at the origin), but it can be removed by
  subtracting the orbitally induced Doppler velocity of the primary from the
  phase-resolved input spectra.

  We also applied our new projection technique to real data of the dwarf nova
  system IP Peg during an outburst which shows emission from the irradiated face
  of the secondary and a distinct spiral structure in the accretion disc.
  The secondary is now placed `outside' the correctly orientated disc in
  velocity space.
  The dominant spiral structure of the shocks in the disc is well preserved and
  the shocks are now spiralling `inwards' from lower to higher velocities.
  Due to the relative high mass ratio ($q\sim0.5$) of IP Peg the vertical
  asymmetry in the brightness distribution in the tomograms with the C.O.M.~or
  anti-C.O.M.~of the binary at the origin is more pronounced than in the case of
  WZ Sge.
  This implies that it is crucial to consider the mass ratio ($q$) of the system
  when interpreting the brightness distribution in a tomogram.
  The asymmetry in the brightness distribution is removed in tomograms with the
  anti-C.O.M.~of the primary at the origin resulting in a more circularly
  symmetric brightness distribution.
  This, together with the `inwards' spiralling appearance of the shocks, creates
  a velocity image which is perhaps more intuitive with the expected radial disc
  profile of the shocks.

  We conclude that our new technique of inside-out Doppler tomography is
  complementary to the existing technique.
  The standard projection used in Doppler tomography tends to concentrate and
  enhance lower velocity features while higher velocity features are more
  separated and dispersed.
  Conversely, the inside-out velocity projection tends to concentrate and
  enhance higher velocity features while lower velocity features are more
  separated and dispersed.
  This is perhaps more consistent with the emission distribution in
  disc-accreting CVs where the high- and low-velocity emissions are produced in
  smaller and larger fractions of the total system volume, respectively.

\begin{acknowledgements}
  We thank Danny Steeghs for providing the IP Peg data and we are grateful to
  Keith Horne for his helpful comments.
  Additionally we thank the referee for useful comments.

  This material is based upon work supported financially by the National
  Research Foundation (NRF).
  Any opinions, findings and conclusions or recommendations expressed in this
  material are those of the author(s) and therefore the NRF does not accept any
  liability in regard thereto.
\end{acknowledgements}

\bibliographystyle{aa} 
\bibliography{refs}    

\end{document}